\documentclass[stats,article,accept,oneauthor,pdftex]{Definitions/mdpi}

\firstpage{1}
\pubvolume{1}
\issuenum{1}
\articlenumber{0}
\pubyear{2022}
\copyrightyear{2020}
\datereceived{Nov 17, 2021}
\dateaccepted{Jan 29, 2022}
\datepublished{}
\hreflink{https://doi.org/} % If needed use \linebreak
\pdfoutput=1

\let\DS=\displaystyle
\let\ga=\gtrsim
\let\la=\lesssim

\newcommand{\eq}[1]{\begin{equation} #1 \end{equation}}

\newcommand{\cE}[1]{\hbox{$\cdot10^{#1}$}}

\newcommand{\half}{\hbox{$\tfrac12$}}

\newcommand{\gu}{\hbox{$g_{\rm u}$}}
\newcommand{\Tu}{\hbox{$T_{\rm u}$}}
\newcommand{\xh}{\hbox{$x_{\rm h}$}}
\renewcommand{\th}{\hbox{$t_{\rm h}$}}
\newcommand{\Qh}{\hbox{$Q_{\rm h}$}}
\newcommand{\qh}{\hbox{$q_{\rm h}$}}
\newcommand{\hk}{\hbox{$h_{\rm k}$}}
\newcommand{\hkk}{\hbox{$h_{\rm k}^k$}}

\newcommand{\ak}{\hbox{$\alpha_k$}}
\newcommand{\Qhat}{\hbox{$\hat Q$}}
\newcommand{\Qbar}{\hbox{$\bar Q$}}
\newcommand{\cov}{\hbox{COVID-19}}
\newcommand{\Fc}{\hbox{$F_{\rm crit}$}}

\graphicspath{{Figs/}}

\newcommand{\new}{\color{red}}
\let\new=\relax

\Title{A General Description of Growth {\new Trends}}

\TitleCitation{A General Description of Growth Trends}

\Author{Moshe Elitzur\orcidA{}}

\AuthorNames{Moshe Elitzur}

\AuthorCitation{Elitzur, M.}

\address{%
    Department of Astronomy,
    Univ of California, Berkeley,
    CA 94720, USA;
    moshe@g.uky.edu}

\corres{Correspondence: moshe@g.uky.com}

\abstract{ Time series that display periodicity can be described with a Fourier
    expansion. In a similar vein, a recently developed formalism enables
    description of growth patterns with the optimal number of parameters
    \cite{EKZ20}. The method has been applied to the growth of national GDP,
    population and the COVID-19 pandemic; in all cases the deviations of
    long-term growth patterns from pure exponential required no more than two
    additional parameters, mostly only one. {\new Here I utilize the new
    framework to develop a unified formulation for all functions that describe
    growth deceleration, wherein the growth rate decreases with time. The
    result offers the prospects for a new general tool for trend removal in
    time-series analysis.
    }}



\begin{document}
%%%%%%%%%%%%%%%%%%%%%%%%%%%%%%%%%%%%%%%%%%

\section{Introduction}
\label{sec:intro}

{\new
Analysis that seeks to identify causal links among components of dynamic
systems requires models that account for all the relevant processes and
interactions that affect the quantities of interest. The complexity of such
models increases rapidly with the complexity of the underlying system and the
forecasting range. An instructive example is Gross Domestic Product (GDP), a
time series with an extremely complex dependency on many national and
international variables. The current UK Treasury Model used for econometric
forecasting utilizes around 30 main equations and 100 independent input
variables \cite{StatsRef21}. Such complexity is unavoidable when modeling is
aimed at understanding the drivers of the time variation.

A simpler approach, employed when the goal is limited to forecasting without
attempting to uncover causal connections, is to describe the data with
statistical indicators that can be calculated with general purpose,
off-the-shelf software packages without regard to the nature of the phenomenon
under study. Apart from rudimentary statistical indicators such as mean and
variance, useful information about a dataset is obtained from time-series
analysis of its stochastic variability. Such analysis utilizes autoregression
(AR) or moving average (MA) process modeling, or a combination of the two
(ARMA) \cite{StatsRef21, Ivezic14}. The underlying assumption is that the time
series is stationary, meaning that the origin of time does not affect the
properties of the studied process. This assumption implies that prior to the
application of stochastic analysis, all systematic components that have
consistency or recurrence must be removed from the time series. A seasonal
component is removed when the series exhibits regular fluctuations based on the
time of the year. Seasonality is always of a fixed and known period. An
additional type of deterministic recurring variation is usually referred to as
cyclical, corresponding to variations that are periodic but not seasonal, or
regular but not of fixed period. In practice, the difference between the two
categories is one of semantics rather than substance --- both can be removed
with Fourier analysis, with seasonal variations described by a single frequency
while cyclic ones require a finite number ($> 1$) of Fourier components. The
removal of all regularly recurring variations from a time series is possible
because Fourier analysis can describe every variation pattern that displays
periodicity of any kind.

Sufficiently long time series can be averaged over multiple segments, each
having a span longer than the longest regularly recurring variation. When a
monotonic trend exists in the resulting sequence of mean values, it implies a
secular variation that cannot be modeled with a combination of Fourier
components. Instead, removal of such monotonic trend, sometimes called
``detrending,'' is commonly done by differencing, leading to autoregressive
integrated moving average (ARIMA) modeling \cite{StatsRef21, Forecasting21,
Ivezic14}. Differencing $n$ times will remove a monotonic trend that varies as
a power-law with index $n$, thus it can remove all polynomial trends. However,
differencing is ineffective for the exponential growth that typifies the
long-term behavior of, for instance, many nations' GDP and population.
Exponential trends can be handled by switching to the logarithm of the data
points, transforming the time series into one with a linear trend that is then
removed by differencing. But although effective, this technique is only
applicable to growth at a constant rate. In particular, it cannot handle
declining growth rates, which are quite common. Such slowdown of growth is
sometimes described with the logistic function (see \S\ref{sec:logist}), which
serves as the basis for logistic detrending \cite{StatsRef21}. But this is
again a specific function with limited applicability. Even in fields where the
logistic had notable successes, such as diffusion of innovation
\cite{Griliches57}, its symmetric S-shape conflicts with much of the data
\cite{Lekvall73} (see \S\ref{sec:discussion}).

In contrast with the handling of periodic variations, where Fourier analysis
provides the foundation for a universal technique, a general method for the
removal of monotonic long-term trends from time series is not yet available.
The prospects for such a framework now exist thanks to the newly developed
\emph{hindering formalism} to extract the exponential component from a growth
process and describe the remainder with the optimal number of parameters
\cite{EKZ20}.
}%
Based on a general solution of the equation of growth, the method has been used
to analyze the time variation of population and GDP in the US and UK, the
countries with the longest continuous datasets, going back more than 200 years.
The results show that in spite of highly volatile growth rates, the long-term
time variations of both GDP and population in both the US and UK are rather
smooth and regular. The formalism has also been used to model the \cov\
pandemic outburst in 89 nations and US states \cite{Elitzur21}. The sizeable
sample enabled a meaningful search for correlations that yielded strong
statistical evidence for the impact of preventive policies on slowing the
pandemic initial growth; a delay of one week in the implementation of the first
policy nearly tripled the size of the infected population, on average.

{\new%
The aim of this paper is to solidify the methodology of the hindering formalism
so that it can become a standard detrending tool in time series analysis. After
deriving the general solution of the equation of growth in \S\ref{sec:growth},
in \S\ref{sec:decelerated} I develop a new general formulation for any function
that describes decelerated growth, including the logistic, and present detailed
analysis and comparisons of these functions. Such meaningful comparison is made
possible thanks to the newly derived unified functional form that uses a common
set of parameters to describe every possible pattern of growth deceleration.
Section \ref{sec:time series} discusses the practical details of implementing
the hindering formalism in data analysis of a time series, and
\S\ref{sec:examples} presents actual examples of such analyses. Accelerated
growth is discussed briefly in \S\ref{sec:accelerated}, which shows that growth
acceleration can only have a limited duration. Section \ref{sec:discussion}
closes with a detailed discussion, including both advantages and limitations of
the formalism presented here and directions for future work.
}

%\newpage

\section{The Equation of Growth and its Solution}
\label{sec:growth}

The growth of quantity $Q\ (> 0)$ with time $t$ is described by the equation of
growth

\eq{\label{eq:growth}
   \frac{dQ}{dt} = gQ
}
where $g$ is the growth rate.\footnote{The growth rate is frequently denoted
$r$ rather than $g$. Here I follow the common practice among economists.} For
this equation to be meaningful it must be accompanied by some suitable
constraints on $g$. A growth process is characterized by a monotonically
increasing $Q$, so $g$ must be positive. This also implies that $Q$ is a
single-valued function of time, therefore $g$ can be considered a function of
$Q$, itself a function of $t$. As a linear differential equation, the solution
requires a boundary condition such as the value of $Q$, say $Q_i$, at some
initial time. The initial $Q_i$ can be arbitrarily small (though $> 0$,
otherwise $Q$ will remain 0 at all times). In addition, $g(Q_i)$, too, must be
$> 0$ (to avoid $gQ = 0$) no matter how small $Q_i$. Therefore the limit $g(Q
\to 0)$ must exist and we require it to be finite\footnote{Appendix
\ref{app:Gompertz} shows an example with $g(Q) \to \infty$ when $Q \to 0$. Such
divergence is excluded from our discussion.} and non-vanishing:

\eq{\label{eq:gu}
   \gu \equiv \lim_{Q \to 0} g(Q)  > 0.
}
We refer to \gu\ as the \emph{unhindered growth rate} for reasons that will
become clear below. Now make the transformation from $g(Q)$ to the function
$f(Q)$, defined from

\eq{\label{eq:f}
   g(Q) = \frac{\gu}{1 + f(Q)}.
}
This transformation effects a complete separation of the variables $t$ and $Q$
in eq.\ \ref{eq:growth}. While $g$ is rate, with dimensions of inverse time,
$f$ is a dimensionless mathematical function. The requirement $g > 0$ implies
$f(Q) > -1$ for all $Q$, and the condition in eq. \ref{eq:gu} translates into
$f(0) = 0$; other than that, $f$ is arbitrary. Assuming it to be a well behaved
function, $f$ can be expanded in a power series

\eq{\label{eq:f_series}
    f(Q) = \sum_{k \ge 1} \ak Q^k,
}
with \ak\ some expansion coefficients; the condition $f(0) = 0$ dictates
$\alpha_0 = 0$. Inserting this series expansion into the growth equation yields

\eq{\label{eq:solution}
   \ln Q + \sum_{k \ge 1} \frac1k\ak Q^k = \gu t + C,
}
where $C$ is a constant determined from the initial condition. This is the
general solution of the equation of growth \cite{EKZ20}. Any growth pattern can
be described by this equation with a suitable choice of the expansion
parameters \ak.

In addition to enabling solution of the growth equation, the transformation
from $g$ to $f$ (eq.\ \ref{eq:f}) also provides a useful classification of the
domains of growth. The point $f = -1$ yields a singularity for $g$, separating
contraction ($g < 0$) from expansion ($g > 0$). The solution in eq.\
\ref{eq:solution} cannot be extended across this singularity, it is
inapplicable to declining quantities. Since the constraint in eq.\ \ref{eq:gu}
cannot be met for a decreasing $Q$, a general description of the $g < 0$ domain
would require a different approach. The growth domain, $f > -1$, is further
divided into two distinct regions by the point $f = 0$, which corresponds to a
simple exponential with the constant growth rate \gu. In the region $-1 < f <
0$ the growth rate obeys $g(Q) > \gu$, corresponding to accelerated growth
--- as $Q$ increases so does the growth rate. The domain $f > 0$ corresponds to
decelerated growth --- $g(Q) < \gu$, growth is slowing as $Q$ is increasing. We
discuss first the latter case, which is more common.

\section{Decelerated Growth}
\label{sec:decelerated}

The equation of growth (eq.\ \ref{eq:growth}) contains two independent units of
measurement, one each for $Q$ (e.g., currency, size of population, etc.) and
$t$ (day, year, etc.). As a result, the solution in eq.\ \ref{eq:solution} is
not suitable for a general analysis of growth patterns because every expansion
coefficient \ak\ has its own dimension (inverse of the unit for $Q$, raised to
the $k$th power). If $Q$ describes GDP, for example, changing the currency unit
will require each \ak\ to be scaled by a different factor, resulting in an
entirely different set of expansion coefficients. For a general classification
of growth patterns, intrinsic scales must be removed so that all quantities are
transformed into dimensionless mathematical variables. Since the growth rate is
measured in units of inverse time, the unhindered growth rate \gu\ (eq.\
\ref{eq:gu}) defines an intrinsic scale for time. The natural independent
variable of the problem is the dimensionless

\eq{\label{eq:x}
   x = \gu t.
}

\subsection{Hindering}
\label{sec:hindering}

To identify a similar intrinsic scale for $Q$ we start with a simple
illustrative example. Consider a desolate island into which apple seeds are
introduced. Some seeds will sprout, apple trees will produce new seeds and the
tree population will grow. The growth rate of the first generation of trees is
\gu\ (eq.\ \ref{eq:gu}), determined by the island's climate, ground fertility,
etc. This rate is maintained so long as individual trees do not interfere with
the growth of each other. Once the number of trees has grown to the point that
tree crowding becomes a significant factor, the growth rate begins to decline
from its initial value, an effect termed \emph{hindering} \cite{EKZ20}: the
growing quantity hinders its own growth when it is sufficiently large. The size
of the tree population at the onset of hindering is a characteristic of the
growth process.

For a general discussion we turn to the transformation in eq.\ \ref{eq:f}.
Growth deceleration implies that $g(Q)$ is decreasing with $Q$, therefore its
mathematical transform $f(Q)$ is monotonically \emph{increasing} from its
initial $f(0) = 0$. The increasing $f$ delineates two domains of growth. As
long as $f(Q) \ll 1$ the growth rate is roughly constant, $g(Q) \simeq \gu$,
and $Q$ grows as an unhindered exponential irrespective of the functional form
of $f$. On the other hand, when $f(Q) \gg 1$ the growth rate becomes $g(Q)
\simeq \gu/f(Q)$. This is the \emph{hindered growth} domain: the growth rate
decreases monotonically from the maximal \gu\ with a time variation controlled
by the specific functional form of $f(Q)$. Varying this form yields growth
patterns that can be very different from exponential.

Introduce the \emph{hindering parameter} \Qh, the magnitude of $Q$ at the point
where $f = 1$; that is, \Qh\ is defined from

\eq{\label{eq:Qh}
   f(\Qh) = 1, \quad \text{i.e.,} \quad
   g(\Qh) = \half\gu.
}
The hindering parameter is an intrinsic property of the growth process, marking
the transition between unhindered growth at $Q < \Qh$ ($f < 1$) and hindered
growth at $Q > \Qh$ ($f > 1$). Denote by \xh\ the magnitude of the independent
variable when $Q = \Qh$, {\new namely, \xh\ is defined from

\eq{\label{eq:xh}
   Q(\xh) = \Qh\,;
}
the corresponding time is $\th = \xh/\gu$. The time variation of $Q$ can be
written in terms of a mathematical hindering function $h$ such that

\begin{samepage}
\eq{\label{eq:math}
  Q(t) = \Qh\,h(\gu t - \xh)
}
where
\[
      h(0) = 1, \qquad h'(0) = \half
\]
\end{samepage}%
and where the prime denotes derivative with respect to $x$; the boundary
condition $h'(0) = \half$ arises from the definition of \Qh\ in eq.\
\ref{eq:Qh}. Inserting this form of $Q$ into the equation of growth (eq.\
\ref{eq:growth}) and following the subsequent steps, the hindering function
$h(x)$ describing the growth process obeys

\begin{samepage}
\eq{\label{eq:h_solution}
   \ln h(x) + \sum_{k \ge 1}\frac1k a_k\left[h^k(x) - 1 \right] = x
}
with
\[
    \sum_{k \ge 1} a_k = 1,
\]
\end{samepage}%
a constraint that follows directly from the boundary conditions in eq.\
\ref{eq:math}. The numerical constants $a_k$ are weight factors intrinsic to
the growth process; the dimensional expansion coefficients in eq.\
\ref{eq:solution} are $\ak = a_k/Q_h^k$. Note that the hindering function
$h(x)$ is the solution of the differential equation

\eq{
    \frac{dh}{dx} = \frac{h}{\DS 1 + \sum_{k \ge 1} a_k h^k}
}
with the boundary condition $h(0) = 1$.

We have derived a \emph{universal representation for decelerated growth}. Every
process of decelerated growth can be described with eq.\ \ref{eq:math}. It is
characterized by a mathematical hindering function $h(x)$, defined in eq.\
\ref{eq:h_solution} by its weight coefficients $a_k$, and by the common set of
parameters \gu\ (eq.\ \ref{eq:gu}), \Qh\ (eq.\ \ref{eq:Qh}) and \xh\ (eq.\
\ref{eq:xh}).
}%
The point \hbox{($x$ = 0, $h$ = 1)} marks the transition from unhindered to
hindered growth. The unhindered domain, $x < 0$, is where $h < 1$ ($Q < \Qh$)
and the logarithmic term dominates the left-hand side of eq.\
\ref{eq:h_solution}, yielding exponential growth. The hindered domain, $x > 0$,
has $h > 1$ ($Q > \Qh$). As a result, the power-law expansion terms dominate
and the logarithm can be neglected. We proceed now to some specific examples of
hindering functions $h(x)$.

\subsection{Single-Term Hindering}
\label{sec:sth}

The simplest hindering functions are obtained when all but one of the hindering
coefficients in eq.\ \ref{eq:h_solution} vanish; from the corresponding
constraint, that coefficient must be unity. Then the growth pattern becomes $h
= \hk(x)$, where the single-term hindering function (sth hereafter) of order
$k\ (\ge 1)$ is defined via

\eq{\label{eq:sth}
   \ln\hk(x) + \frac1k\left[\hkk(x) - 1\right] = x.
}
This is an implicit analytic definition of \hk. For any given $x$, $\hk(x)$ can
be calculated numerically from this equation with a suitable procedure; the
Newton method proved to be both efficient and reliable. The time variation of
the associated growth rate is

\eq{\label{eq:sth_g}
   g(\hk) = \frac{\gu}{1 + \hkk(x)}.
}
All sth functions have $\hk(0) = 1$. Leading-order approximation for the
behavior of \hk\ when $x < 0$ ($\hk < 1$) are obtained by neglecting \hkk\ and
retaining only $\ln\hk$ in eq.\ \ref{eq:sth}, with the opposite approximation
when $x
> 0$ ($\hk > 1$). This yields

%https://tex.stackexchange.com/questions/240868/how-to-write-cases-with-latex
\eq{\label{eq:sth_approx}
    \hk(x) \simeq \left\{%
    \begin{array}{ll}
        e^{x + 1/k}     &\qquad x \ll 0\\ \\
        (1 + kx)^{1/k}  &\qquad x \gg 0
    \end{array}\right.
}
In the unhindered domain \hk\ increases exponentially for all $k$. In the
hindered domain its asymptotic behavior is $\hk \propto x^{1/k}$; the larger is
$k$, the slower the growth.

%%%%%%%%%%%%%%%%%%%%%%%%%%%%%%%%%%%%%%%%%%%%%%%%%%%%%%%%%%%%%%%%%%%%%%%%%%
\begin{figure}[ht]%[H]
\includegraphics[width=\hsize, clip]{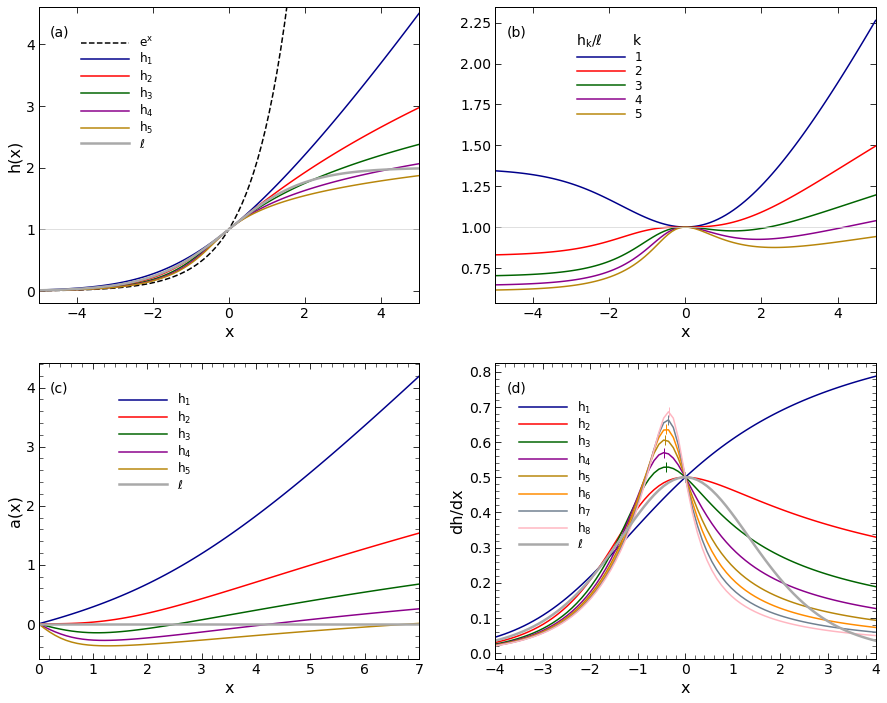}
\centering \caption{Mathematical hindering functions. (\textbf{a}) Plots of
single-term hindering \hk\ (eq.\ \ref{eq:sth}) for various values of $k$, as
labeled, and of the logistic $\ell$ (eq.\ \ref{eq:logist}). A thin horizontal
line at $h = 1$ marks the boundary between the unhindered ($x < 0,\ h < 1$) and
hindered ($x > 0,\ h > 1$) domains. Also shown is the exponential $e^x$ in
dashed line. (\textbf{b}) The ratio $\hk(x)/\ell(x)$. (\textbf{c}) The
asymmetry measure $a$ (eq.\ \ref{eq:asymmetry}) for \hk\ and $\ell$.
(\textbf{d}) Time derivatives of \hk\ and $\ell$ (eq.\ \ref{eq:h'}). All pass
through the point $\left(0, \half\right)$ (see eq.\ \ref{eq:math}). Except for
$h_1$, every derivative has a peak, marked with a short vertical line.
}
\label{fig:Mosaic}
\end{figure}
%%%%%%%%%%%%%%%%%%%%%%%%%%%%%%%%%%%%%%%%%%%%%%%%%%%%%%%%%%%%%%%%%%%%%%%%%%

Panel (\textbf{a}) of Figure \ref{fig:Mosaic} shows plots of \hk\ for $k =
1,\ldots, 5$. For comparison, the exponential function is also shown, plotted
with a dashed line. The sth functions are defined only for $k \ge 1$ (eq.
\ref{eq:sth}). However, $(z^k - 1)/k \to 0$ when $k \to 0$ for any value of
$z,$\footnote{This is easily verified with L'Hôpital's rule.} therefore we can
formally consider $e^x$ as the 0-th order member of the \hk\ series, consistent
with the $k \to 0$ limit in the definition of \hk.

\subsection{Multi-Term Hindering}

Every sth function increases without a bound when $x \to \infty$, although the
rise flattens with increasing $k$. When the hindering sum in eq.\
\ref{eq:h_solution} is dominated by its $k$th order term, the asymptotic
behavior of $h$ is $\sim x^{1/k}$ (eq.\ \ref{eq:sth_approx}) --- larger values
of $k$ provide flatter growth. Therefore, when the sum contains a finite number
of terms with monotonically decreasing $a_k$, $h$ varies as follows: After an
initial exponential rise, the linear term in the sum starts to dominate when
$a_1h$ \hbox{becomes $> 1$}, and $h$ becomes proportional to $x$ instead of
$e^x$. Once the 2nd-order term starts dominating, the behavior switches to
$h~\propto~x^{1/2}$, then flattens further to $h~\propto x^{1/3}$ and so on.
Finally, when the sum's last term, with $k = k_{\rm max}$, dominates, the time
variation settles into $h~\propto~x^{1/k_{\rm max}}$, unbounded growth that
continues indefinitely. \emph{A finite sum of hindering terms describes
unbounded growth}. In the limit $k_{\rm max} \to \infty$, the $x^{1/k_{\rm
max}}$ behavior approaches a constant that sets an upper limit on $Q$.
\emph{Bounded growth requires hindering series with infinite numbers of terms}.
We now describe one particular example of bounded growth.

\subsection{The Logistic}
\label{sec:logist}

The logistic growth function is employed in many fields, including population
studies \cite{Pearl24, Schacht80, Kingsland82}, diffusion of technology
\cite{Griliches57}, natural selection \cite{Pianka70} and GDP growth
\cite{KWASNICKI13}. Its underlying mathematical function normalized to unity at
$x = 0$ is $h = \ell(x)$, where

\eq{\label{eq:logist}
 \ell(x) =  \frac{2}{1 + e^{-x}}.
}
At large $x$ the function approaches the limit $\ell(x \to \infty) = 2$, thus a
quantity $Q$ varying as the logistic has the upper bound $K = 2\Qh$ (eq.
\ref{eq:math}), called the \emph{carrying capacity}. The approximate behavior
in the unhindered and hindered domains is

\eq{\label{eq:logist_approx}
    \ell(x) \simeq 2\times\left\{%
    \begin{array}{ll}
       e^x             & \qquad x \ll 0    \\ \\
       (1 - e^{-x})    & \qquad x \gg 0
    \end{array}\right.
}
As with all growth functions, the logistic increases exponentially in the
unhindered domain. In the hindered domain it approaches rapidly the limit of 2;
at $x = 3$, $\ell(x)$ is already within 5\% of its upper bound. The logistic
growth rate is\footnote{Inserting $\ell(x)$ from eq.\ \ref{eq:logist}, the
growth rate as a function of time is $g(x) = \gu/(1 + e^x)$, implying that
$f(x) = e^x$.}

\eq{\label{eq:logist_g}
   g(\ell) = \gu\left[1 - \half\ell(x)\right].
}
It vanishes as $Q$ reaches the carrying capacity. From eq.\ \ref{eq:f}, the
associated $f$-transform is

\eq{
   f(\ell) = \frac{1}{1 - \half\ell(x)} - 1
        = \sum_{k = 1}^{\infty} \left(\half\ell\right)^k.
}
As expected for bounded growth, the logistic hindering series (eq.\
\ref{eq:h_solution}) is infinite, with expansion coefficients $a_k = 1/2^k$.

\subsection{Comparison, logistic vs sth}
\label{sec:comparison}

In addition to sth functions, panel (\textbf{a}) of Figure \ref{fig:Mosaic}
shows also a plot of the logistic, which stands out with its distinct S-shape.
As a bounded-growth function, the logistic is overtaken by every sth function,
although $x$ at the overtake point increases with $k$. The differences between
the sth functions and the logistic are accentuated in panel (\textbf{b}), which
shows their ratios. At negative $x$, the ratio $\hk(x)/\ell(x)$ is
approximately $\half e^{1/k}$ (see equations \ref{eq:sth_approx} and
\ref{eq:logist_approx}). In particular, when $x \ll 0$ and $k = 1$ the ratio
approaches $\half e = 1.36$, while for $k = 2$ it is $\half\sqrt{e} = .824$; as
$k$ increases, the ratio approaches \half. At positive $x$, the ratio increases
without bound.

The logistic S-shape obeys $\ell(x) - 1 = 1 - \ell(-x)$, a reflection symmetry
about $(0, 1)$.  There is no similar symmetry relation for the sth functions,
which vary roughly exponentially to the left of this point and as a power to
the right of it (eq.\ \ref{eq:sth_approx}). Panel (\textbf{c}) of Figure
\ref{fig:Mosaic} shows the asymmetry measure of the various hindering
functions, defined as

\eq{\label{eq:asymmetry}
    a(x) = \frac{h(x) - 1}{1 - h(-x)} - 1.
}
For the logistic, $a(x)$ is identically 0. For sth, the asymmetry increases
without a bound when $x \ga 1$.

The time derivatives of the sth and logistic functions, shown in panel
(\textbf{d}) of Figure \ref{fig:Mosaic}, are

\eq{\label{eq:h'}
 \frac{dh}{dx} = h\times\left\{%
    \begin{array}{ll}
       \frac{\DS 1}{\DS1 + h^k}  & \qquad \text{sth}       \\ \\
       (1 - \half h)             & \qquad \text{logistic}
    \end{array}\right.
}
All functions have $h' \to 0$ when $x \to -\infty$ (i.e., $h \to 0$), and $h'$
also vanishes when $x \to \infty$ for every function except for $k = 1$ sth. As
a result, $h'$ peaks at some finite $x$ for all functions other than $k = 1$
sth, whose derivative increases monotonically toward an upper limit of 1. The
derivative peaks are marked with short vertical lines in panel
(\textbf{d}).\footnote{The logistic peak derivative is $\ell' = \half$ at $x =
0$. The peak derivative of \hk\ for $k > 1$ is $k^{-1/k}\left(1 -
\frac1k\right)^{1 - 1/k}$ at $x = -\frac1k\left[\ln(k - 1) + \frac{k - 2}{k -
1} \right]$.}  The peak of $h'_2$ is \half\ at $x = 0$, same as the logistic.
As $k$ increases, the peak location first moves to the left, then back toward
$x = 0$; the leftmost peak is at $x = -0.441$ when $k = 4$. The peak value of
$h'_k$ is slowly approaching unity as $k \to \infty$.

\section{Handling of Time Series}
\label{sec:time series}

A time series is a sequence of measurements $Q_0, Q_1, \dotso$ taken at
monotonically increasing times $t_0 < t_1 < \dotsc$; without loss of
generality, $t_0$ can be taken as 0. The time intervals are frequently equal to
each other, but this is not a requirement.

The series describes a growth process if it displays an overall trend of
monotonic increase. The key here is long-term behavior --- a time series of
national GDP, for example, may contain segments of decline during occasional
recessions but still maintain an overall trend of growth. The presence, or
absence, of a monotonic trend can be conveniently determined with the
Mann-Kendall trend test (hereafter MK test), commonly employed  in studies of
environmental, climatological and hydrological data \cite{Kocsis17}. The test
involves the sum

\eq{
    S = \sum_{i}\sum_{j>i} \text{sgn}(Q_j - Q_i),
}
where sgn($x$), the sign function, is 0 if $x = 0$ and $|x|/x$ otherwise. The
test's null hypothesis (H$_0$) is no trend in the time series. In that case the
MK statistic $Z$, obtained from $S$ through normalization by the expected
variance, follows the normal distribution with a zero mean and unity standard
deviation. Positive (negative) $Z$ indicates an increasing (decreasing) trend;
for example, $Z = 3$ is a 3$\sigma$ evidence for a growth trend. This
non-parametric test can detect a monotonic trend in time series of at least 8
members \cite{Blain13} without assuming the data to be distributed according to
any specific rule (in particular, there is no requirement of normal
distribution).

Given a time series, we first determine whether it describes a growth process
by testing the MK null hypothesis against the alternative hypothesis (H$_{\rm
a}$) that there is an increasing monotonic trend ($Z > 0$) in a one-tailed
test. When a long-term growth trend is detected, the next step is to test for
the presence of growth slowdown. For that we compute the time series of growth
rates $g_0, g_1, \dotsc$ from a finite-difference calculation of the pairs
$(Q_0, t_0), (Q_1, t_1)\dotso$ and MK-test this series for a \emph{decreasing}
trend ($Z < 0$). When the dataset does correspond to a growth process with a
decreasing growth rate it can be described by a hindering function with the aid
of eqs.\ \ref{eq:math} and \ref{eq:h_solution}. The shift of independent
variable from the (inherently arbitrary) time origin $x = 0$ is $\xh =
-h^{-1}(Q_0/\Qh)$, where $h^{-1}$ is the inverse of the pertinent hindering
function. For the functions considered above (\S\S\ref{sec:sth},
\ref{sec:logist}) these shifts are

\eq{
  \xh = \left\{%
  \begin{array}{ll}
     \ln\qh + \DS\frac{1}{k}(1 - q_{\rm h}^{-k})  &\qquad \text{sth} \\ \\
     \ln(2\qh - 1)                                &\qquad \text{logistic}
  \end{array}\right.
}
where $\qh = \Qh/Q_0$. When $\qh > 1$, the logistic reaches hindering before $k
= 1$ sth; as $k$ increases, sth reaches hindering first, with \gu\th\
decreasing toward $\ln\qh$.

%%%%%%%%%%%%%%%%%%%%%%%%%%%%%%%%%%%%%%%%%%%%%%%%%%%%%%%%%%%%%%%%%%%%%%%%%%
\begin{figure}[ht] %[H]
\includegraphics[width=\hsize, clip]{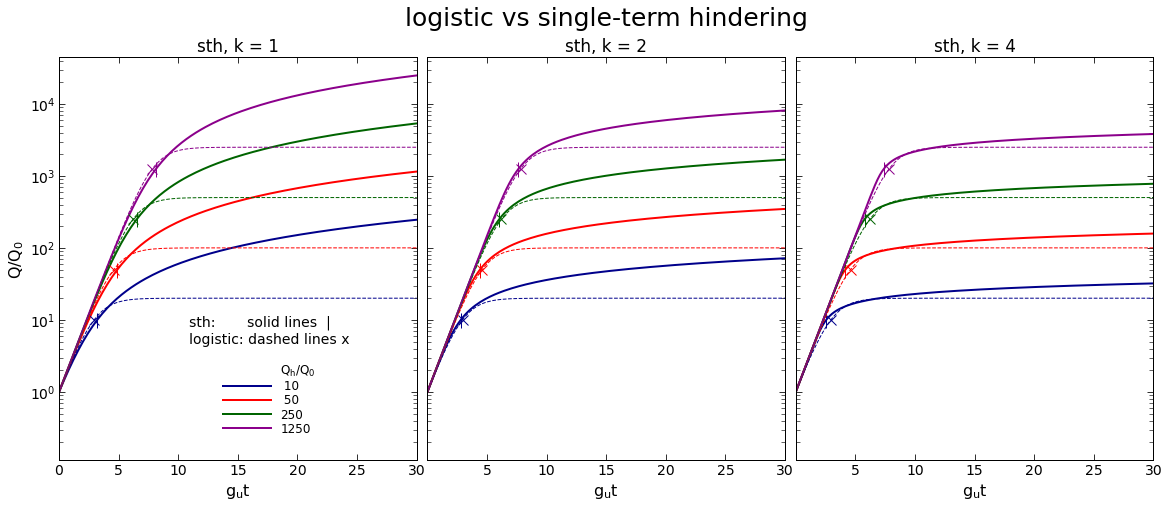}
\centering \caption{Comparison of growth processes $Q$ (eq.\ \ref{eq:math})
that follow the sth (eq.\ \ref{eq:sth}) and logistic (eq.\ \ref{eq:logist})
functions with the same \gu, $Q_0$ and \Qh\ for various values of $\Qh/Q_0$, as
marked. Solid lines show sth functions, with short vertical marks ($\vert$) at
$Q = \Qh$; the logistic is plotted with dashed lines, the hindering marker is
$\times$. Each panel shows sth with a different $k$, as labeled. The logistic
curves are the same in all panels.
}
\label{fig:logist_hind}
\end{figure}

%%%%%%%%%%%%%%%%%%%%%%%%%%%%%%%%%%%%%%%%%

{\new%
Thanks to the unified formulation of decelerated growth functions, we can now
compare different hindered growth patterns described by the same common set of
parameters.
}
Figure \ref{fig:logist_hind} shows plots of $Q$ for sth and logistic functions
that have the same \gu, $Q_0$ and \Qh. Each plot is obtained from the
corresponding mathematical function shown in panel (\textbf{a}) of Figure
\ref{fig:Mosaic} by shifting the $x$-axis origin and scaling the $y$-axis as
prescribed in eq.\ \ref{eq:math}. On each plot, the hindering point $Q = \Qh$
is marked. To its left is the unhindered growth domain with the universal
$e^{g_ut}$ behavior; the larger is \qh, the longer the exponential rise. To the
right is the hindered growth domain, displaying the differences between the sth
and logistic functions discussed in \S\ref{sec:comparison}.

\subsection{Fitting procedures}
\label{sec:fitting}

When the members $Q_i$ of a time series display decelerated growth we calculate
model points $\Qhat_i = Q(t_i)$ according to equations \ref{eq:math} and
\ref{eq:h_solution}. The best-fitting model parameters are obtained by
minimizing the residual sum of squares (RSS) of the data and model points.
Because of the large dynamic range spanned by typical datasets, we give all
data points equal relative weights ($\sigma_i \propto Q_i$) so that the
minimization is performed on $\text{RSS} = \sum_i\left(\Qhat_i/Q_i -
1\right)^{\!2}$. It is important to note that we only seek the minimum of RSS;
its actual magnitude is immaterial (no need to specify the proportionality
constant in $\sigma_i \propto Q_i$).

Equation \ref{eq:h_solution} is the general solution of the equation of growth
and thus can describe any time series of growth process, given a sufficient
number of expansion coefficients. However, adding terms indiscriminately in
search of a smaller error runs the risk of overfitting and chasing structures
that may reflect noise, not fundamental trends. Our aim, instead, is to
identify the long-term trends in the data rather than construct the absolute
best fit. For that we first model the dataset with a single hindering term and
determine the power $k$ that provides the best fit. The logistic is
parameterized by the same set of variables, \gu, \Qh\ and \xh, and we determine
the best fit with this function too. Between the two resulting fits, the one
with the smaller RSS error is the best minimal hindering model, containing just
one free parameter more than a pure exponential. When the minimal model is
single-term hindering, we proceed to add another term and search for the pair
of power-law indices that yield the best-fitting two-term model (eq.\
\ref{eq:h_solution}). Since the addition of a term will in itself improve
fitting, we must determine the statistical significance of such improvement.
The single-term model is a restricted form of the two-term model, with the
coefficient of the 2nd term restricted to 0, thus the problem can be handled
with the $F$-test, assuming that the unobserved error is normally distributed
\cite{Econometrics09}.\footnote{The $F$-test is closely related to the
\emph{odds ratio} test in Bayesian statistics. The two become the same if and
only if one assumes scale-invariant Jeffreys’ prior for RSS \cite{Ivezic14}.}
The $F$-test null hypothesis is that the additional term has no effect on the
dependent variable so that its coefficient should be 0. The number of data
points, the ratio of RSS for the two models and their number of free parameters
are combined to form the $F$-statistic (or $F$ ratio); it follows an
$F$-distribution, which arises as the ratio of two normal random variates. The
$F$-statistic is compared with a critical value \Fc, determined by the degrees
of freedom for each model and an accepted error level $\alpha$. When $F > \Fc$,
the null hypothesis can be rejected at the confidence level $1 - \alpha$, the
probability of a false rejection is less than $\alpha$. When that is the case,
the improvement from the additional term is statistically meaningful and the
process can be repeated, adding higher terms one-by-one until the improvement
becomes statistically insignificant.

\section{{\new Sample Applications}}
\label{sec:examples}

{\new We now present applications of hindering analysis to actual datasets.
These examples showcase the power and versatility of the new hindering
formalism. While earlier versions of these analyses have already been reported
\cite{EKZ20, Elitzur21}, the formulation in \S\ref{sec:hindering} of a
universal description for decelerated growth provides newly gained insight into
the successes and difficulties of these modeling efforts.}

\subsection{US Population and GDP}
\label{sec:US}

\begin{figure}[ht]%[H]
\includegraphics[width=\hsize, clip]{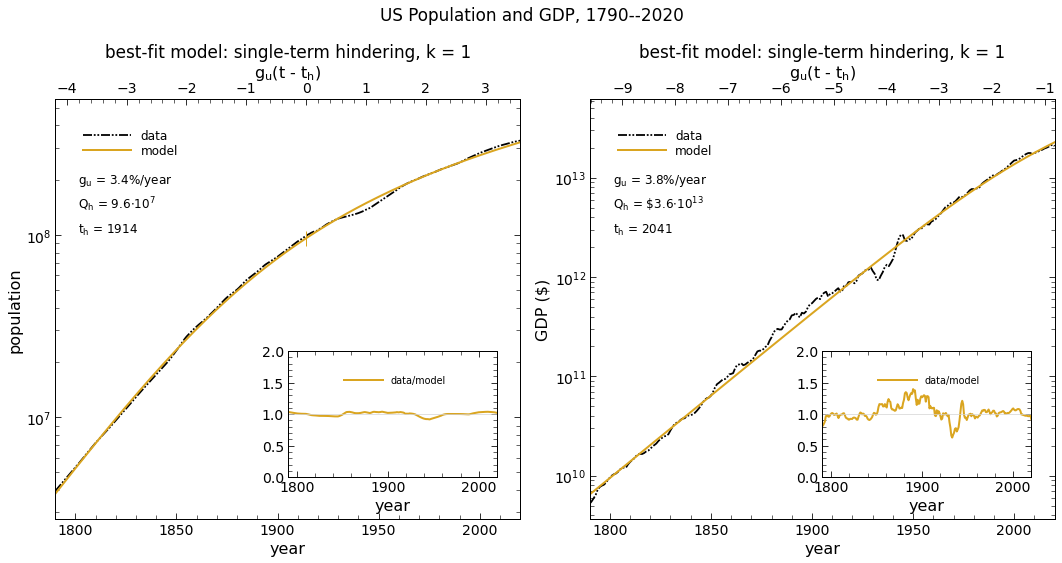}
 \centering
\caption{US Population (\textbf{left}) and GDP (\textbf{right}) from 1790 to
2020. The data are shown in dashed-dotted-dotted line; the GDP currency unit is
\$\ = 2020 USD. In both cases, the best-fitting model, shown in solid line, is
$k = 1$ sth with the listed parameters. The top axis shows $x - \xh$ (cf eqs.\
\ref{eq:x}, \ref{eq:math}). The inset shows the ratio of data to model.
}
\label{fig:US}
\end{figure}

The US and UK are two nations with continuous GDP and population data going
back more than 200 years. Hindering analysis of their data to 2018 was
presented in \cite{EKZ20}. With two more years of data, here we repeat the
analysis of US annual GDP and population data from 1790--2020
\cite{Johnston21}, a total of 231 points for each time series. Although each
dataset contains two additional points, the modeling results, shown in Figure
\ref{fig:US}, are identical to those in \cite{EKZ20}.

%%%%%%%%%%%%%%%%%%%%%%%%%%%%%%%%%%%%%%%%%%%%%%%%%%%%%%%%%%%%%%%%%%%%%%%%%%
%%%%%%%%%%%%%%%%%%%%%%%%%%%%%%%%%%%%%%%%%%%%%%%%%%%%%%%%%%%%%%%%%%%%%%%%%%

The figure's left panel shows modeling of the population data. The best-fitting
model is $k = 1$ sth (linear hindering) with the listed parameters. The model
finds that the hindered domain was entered in 1914, and predicts a 2050
population of 400 million, growing at 0.65\% per year. The best-fitting
logistic provides a greatly inferior fit, with RSS error that is 6 times larger
than for the displayed model; moreover, it has $K = 311$ million, an upper
limit to the US population that was surpassed already in 2010. {\new The
addition of another hindering term makes a negligible impact on the fit;
single-term hindering yields the optimal fit to the data. The ratio data:model,
plotted in the inset, shows that the model properly captures the long-term
variation of the time series. The fraction of variance unexplained (fvu = $1 -
R^2$, where $R^2$ is the coefficient of determination) is 2.07\cE{-3}.} The
prediction for 2020 of the model based on the data to 2018 is only 2\% off the
actual population. Discarding as much as the final 40\% of the time series, the
truncated series model predictions for 2050 are within 10\% of those for the
full dataset.

The right panel of Figure \ref{fig:US} shows analysis of the US GDP data. The
best-fitting model again is linear hindering with the listed parameters. {\new
The fit has fvu = 3.36\cE{-3}. Adding a second term yields a marginal
improvement to the RSS error, which the F-test rejects as statistically
insignificant.} This time the hindering threshold has not yet been crossed; the
model predicts this to happen only in 2041, when the GDP will reach \$36
trillion. It is also much more difficult now to distinguish the $k = 1$ sth
from the logistic. The two functions provide equally adequate fits --- the RSS
error is 3.95 for the former vs 3.98 for the latter. For the year 2050, the
linear hindering model predicts a GDP of \$42 trillion, growing at 1.76\% per
year. The logistic's prediction is a GDP of \$35 trillion growing annually at
1.02\%, ultimately bounded by an upper limit of \$48 trillion.

\subsection{\cov\ Outburst}
\label{sec:covid}

Hindering analysis of the \cov\ pandemic first wave was reported for 89 nations
and US states \cite{Elitzur21}. Here we reproduce the results for the \cov\
case counts in New York State, one of the hardest hit locations in the pandemic
early days.

%%%%%%%%%%%%%%%%%%%%%%%%%%%%%%%%%%%%%%%%%%%%%%%%%%%%%%%%%%%%%%%%%%%%%%%%%%
\begin{figure}[ht]%[H]
\centering
\includegraphics[width=\hsize, clip]{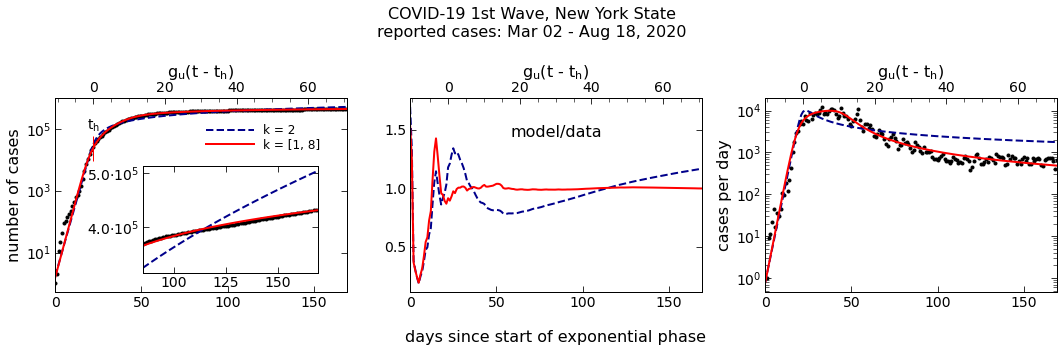}
\caption{The \cov\ pandemic first wave in New York State. Dots show the data,
dashed and solid lines are best fits with eq.\ \ref{eq:h_solution} for single-
and two-term models, respectively. \textbf{Left}: Cumulative number of reported
cases from the start of the exponential outburst; this phase ended in the
transition to hindering, marked \th. The inset zooms in on the second half of
the data with a linear $y$-axis instead of logarithmic. {\new\textbf{Middle}:
Ratio of model to data for each of the fits in the left panel.} \textbf{Right}:
Daily counts. The model curves involve no fitting, being fully determined from
those for the cumulative counts in the left panel.
}
\label{fig:NY_covid}
\end{figure}
%%%%%%%%%%%%%%%%%%%%%%%%%%%%%%%%%%%%%%%%%%%%%%%%%%%%%%%%%%%%%%%%%%%%%%%%%%

The first wave of New York \cov\ cases lasted 170 days, from March 2 to August
18, 2020. Figure \ref{fig:NY_covid} shows the case counts with dots; the left
panel shows the cumulative counts ($Q$), the right one the daily counts
($dQ/dt$). Evident in the left panel is an initial exponential rise followed by
``flattening of the curve,'' corresponding to, respectively, unhindered and
hindered growth. The more moderate growth during the latter phase is better
discerned in the inset, which zooms in on the second half of the dataset with a
linear, instead of logarithmic, $y$-axis.
%{\new
The best fitting minimal hindering model for the cumulative counts (left panel)
is $k = 2$ sth, shown in dashed blue line; its RSS error is 3 times smaller
than the logistic error. The best-fitting two-term model, shown in solid red
line, has $k = [1, 8]$.\footnote{{\new Thanks to improved handling of
higher-order terms, this model is superior to the one presented in
\cite{Elitzur21}.}} Its RSS error is an improvement by factor 1.67 over the
best-fitting single term; the F-test shows this improvement to be statistically
highly significant, with a p-value of 1.11\cE{-16}. While the two models are
hardly distinguishable from the data and from each other on the logarithmic
scale, their differences are evident in the inset {\new and stand out in the
middle panel, which shows the ratio of model to data}. The right panel shows
the model fits to the daily counts. It is important to note that the curves in
this panel involve no fitting; they are fully derived from the models in the
left-panel.

{\new%
The two-term model provides the optimal fit to the data. An additional term
(the best-fitting 3-term model has $k = [1, 2, 9]$) improves the RSS error by
only 0.32\%; the F-test finds this marginal improvement statistically
insignificant with p = 0.47. As is evident from the middle panel, the two-term
model captures the time series long-term trend rather well, with fvu =
3.79\cE{-4}. After some fluctuations around the trend line during the initial
exponential phase, the mean deviation of model from data during the final 117
days (fully 70\% of the time series) is 0.65\%, the maximum just under 2\%.

\subsection{Data Range}

Of the cases presented here, the US GDP stands out as the time series whose
best fit remains ambiguous --- there is no meaningful way to choose between the
logistic and linear hindering ($k = 1$ sth) fits. The underlying cause of the
problem is the range of the independent variable $x$ (eqs.\ \ref{eq:x},
\ref{eq:math}) sampled by the data. The top axis of the GDP plot (right panel
of Figure \ref{fig:US}) shows this range to be [-9.6, -0.8], entirely within
the unhindered domain. As is evident from the top two panels of Figure
\ref{fig:Mosaic}, the logistic and all sth functions are practically
indistinguishable from each other when $x \la -4$ because they all are
proportional to the exponential in that region (eqs.\ \ref{eq:sth_approx},
\ref{eq:logist_approx}). The ratio $h_1(x)/\ell(x)$ is constant to within 1\%
until 1941 (at that year $x = -3.84$). The two functions become distinguishable
afterward, but separate by more than the data fluctuations only around the year
2000. In other words, the entire power to resolve the two fits comes from the
final 20 years of data, which comprise less than 10\% of the time series.
Another 10 data points will add 50\% to the crucial part of the time series. It
can thus be expected that the next ten years or so will enable a selection
between the logistic and linear hindering.

In contrast with the GDP, the linear hindering model for the US population is
decisive, thanks to the propitious range sampled by the data. From the top axis
of the population plot (right panel of Figure \ref{fig:US}), $x$ covers the
range [-4.2, 3.6]. Although the extent of this range is slightly smaller than
for the GDP, the top panels of Figure \ref{fig:Mosaic} show that its placement
provides a clear, unambiguous separation of the $k = 1$ sth function from the
logistic. The NY \cov\ data (Figure \ref{fig:NY_covid}) stand out even further,
with an $x$-range of [-10.7, 70.8], roughly 8 times larger than for the US
population and GDP. This range is so much larger because of the steepness of
the pandemic's initial rise, with \gu\ = 48.2\% per day. Thanks to its large
range, this time series offers a valuable example of the contribution of more
than one hindering term in eq.\ \ref{eq:h_solution}. It is remarkable that two
terms describe so accurately such a large range of $x$.

This discussion highlights the insight provided by the unified description for
all hindering functions (\S\ref{sec:hindering}). The common set of parameters
enables assessment of the significance of derived models and the confidence in
their fits, and helps in making an informed estimate of the range of data
needed for decisive fits.

}

\section{Accelerated Growth}
\label{sec:accelerated}

Accelerated growth, $dg/dQ > 0$, is prone to runaway instabilities. Consider a
small perturbation $\delta Q$ to a random point \Qbar\ in a growth process so
that $Q = \Qbar + \delta Q$. Inserting in the equation of growth (eq.\
\ref{eq:growth}) and retaining only terms to 1st order in $\delta Q$, the
perturbation varies according to

\eq{\label{eq:stability}
        \frac{d\,\delta Q}{dt} =
   \delta Q\times\left(g + Q\frac{dg}{dQ}\right)\Big\rvert_{Q = \bar Q}
}
A small perturbation will decay exponentially when $dg/dQ < -g/Q$ but diverge
exponentially away from the existing pattern whenever $dg/dQ > 0$.
\emph{Accelerated growth is inherently unstable.}

Apart from its inherent instability, the duration of accelerated growth is
limited in general. A simple example of growth acceleration is derived from the
logistic by changing the interaction sign in the growth rate (eq.\
\ref{eq:logist_g}) to give

\eq{
    g(Q) = \gu\left(1 + \frac{Q}{K} \right),
}
a growth rate that increases linearly with $Q$. Here the parameter $K$ denotes
$g(K) = 2\gu$. With the time origin taken at the point where $Q = K$, the
solution of the growth equation is $Q = Ke^x/(2 - e^x)$. Because of the runaway
singularity at $x = \ln2$, the time span of this accelerated growth is limited
to $t < g_{\rm u}^{-1}\ln2$. Now turn to the transformation in eq.\ \ref{eq:f}
for a general description of accelerated growth with control over
singularities. Accelerated growth occurs when $-1 < f < 0$. The lower limit on
$f$ is the transition from growth ($g > 0$) to contraction ($g < 0$), with a
singularity for $g$ at that boundary. The upper limit marks the transition from
a rising $g\ (> \gu)$ to a declining one. With a finite number of expansion
terms for $f(Q)$ (eq.\ \ref{eq:f_series}), the singularity at $f = -1$ is
avoided when the polynomial $1 + f(Q)$ has only imaginary roots. But it is
impossible to simultaneously keep $f < 0$ and prevent an end to growth
acceleration, as illustrated by the polynomial with just $k$ = 1 and 2 which
yields

\eq{
   g(Q) = \frac{\gu}{1
                   - \frac{Q}{K}
                   + \alpha\left(\frac{Q}{K}\right)^{\!\!2}}
}
with $\alpha$ a free parameter. The denominator is the lowest order polynomial
to produce accelerated growth and avoid contraction ($g < 0$); the constraint
$\alpha > \tfrac14$ ensures a positive $g(Q)$ for all $Q$. Growth is
accelerating --- $g(Q)$ increases with $Q$ --- as long as $Q < K/(2\alpha)$.
However, $g(Q)$ reaches a peak of $\alpha\gu/(\alpha - \tfrac14)$ at $Q =
K/(2\alpha)$. Increasing $Q$ further, $g(Q)$ starts to \emph{decrease} ---
growth acceleration turns into deceleration as the quadratic term begins to
dominate. Finally, $g(Q)$ decreases below \gu\ when $Q > K/\alpha$ and the
growth process becomes practically indistinguishable from $k = 2$ single-term
hindering (\S\ref{sec:sth}).

Similar reasoning applies to higher order polynomials, showing that while the
$f = -1$ singularity is avoidable, the switch from accelerated to decelerated
growth at $f = 0$ is not. Growth acceleration cannot be sustained indefinitely.

\section{Discussion}
\label{sec:discussion}

{\new

We developed here a unified scheme for all patterns of decelerated growth
\hbox{(\S\ref{sec:hindering}).} Employing a common set of parameters, this
uniform description enables methodical, systematic selection of the functional
form most suitable for modeling a given dataset. This is especially important
for the handling of growth. While inaccuracies in describing recurring
phenomena are limited by the amplitudes of the variations, there is no bound on
the amount of divergence between different growth trends that are fundamentally
exponential. An instructive example is provided by US population forecasting.
}
In 1924 R. Pearl modeled decadal US census data from 1790--1910 with the
logistic function and concluded that the US population was bounded by an upper
asymptote of 197 million \cite{Pearl24}. In 1966, just 42 years later, this
absolute upper limit was surpassed. Having reached 330 million in 2020, almost
70\% above Pearl's predicted limit, the US population is yet to show signs of
an upper bound. Notably, Pearl's model parameters amounted to \gu\ = 3.13\% per
year, \Qh\ = 98.6 million and \th\ corresponding to the year 1914, nearly
identical to the best-fitting model parameters derived from the 1790--2020 data
in \S\ref{sec:US} (see Figure \ref{fig:US}). The problem with Pearl's
prediction was not the parameters but the fitting function. His model predicts
a 2020 population of 191 million. Using his own parameters but with $k = 1$ sth
instead of the logistic, Pearl would have predicted a 2020 population of 317
million. It is remarkable that a 1924 demographer could have predicted the 2020
US population to within 4\% with just a single-parameter modification to the
exponential function.

Although Pearl missed badly on the US population future growth, his conclusion
was inevitable. The hindering boundary (eq.\ \ref{eq:Qh}) was crossed in 1914,
when the growth rate declined to half its initial, unhindered value, setting
that year's population as \Qh. Having Committed himself to the logistic, Pearl
had to conclude that the carrying capacity was twice \Qh\ (\S\ref{sec:logist}),
hence $K$ = 197 million. Adopting the logistic to model hindered growth implies
an upper limit. Although justified in studies of, e.g., life expectancy
\cite{Marchetti96}, there is no reason why an upper limit should be imposed a
priori on every growth process. When an upper limit does exist, the logistic
dictates it to be 2\Qh\ because of its S-shape symmetry (panel \textbf{a},
Figure \ref{fig:Mosaic}). However, even though diffusion of innovation provides
examples of successful logistic fits \cite{Griliches57}, most diffusion curves
actually show \emph{asymmetric} S-shape, usually the upper shank of the ``S''
is more extended  \cite{Lekvall73}. Such asymmetry implies positive values for
the parameter $a$ (eq. \ref{eq:asymmetry}), shown in panel (\textbf{c}) of
Figure \ref{fig:Mosaic}. Unlike the logistic, every sth function does display
this type of asymmetry, though positive $a$ values start at increasingly larger
$x$ when $k \ge 3$.

The recognition that the logistic is not a universal modeling function even for
bounded growth led to attempts to generalize it with additional parameters
\cite{Pearl24} or combinations of different logistics \cite{Lekvall73}, but
these attempts were based on ad-hoc assumptions. By contrast, the formalism
presented here does not prescribe a priori any specific form for the modeling
function. Instead, the functional form is determined from the data through a
parametrization of the general solution of the equation of growth (eq.\
\ref{eq:h_solution}). Applicable to both bounded and unbounded growth, this
solution provides a generic description of growing quantities just as the
Fourier series provides a generic description of periodic phenomena. All growth
processes share some general properties. The growth of any quantity $Q$ occurs
within some environment, broadly defined as the collection of all the processes
and system components that affect the growth of $Q$ other than $Q$ itself. As
long as the growing $Q$ is sufficiently small that its impact on the
environment is negligible, its growth rate is determined solely by intrinsic
properties of the environment; this is the unhindered growth rate \gu\ defined
in eq.\ \ref{eq:gu}. This rate is maintained until $Q$ becomes sufficiently
large that it significantly impacts the environment, at which point it also
affects its own growth rate. In general, this causes the growth rate to
decline, the effect we refer to as hindering --- the growing quantity has
become so large as to hinder its own growth.

One interpretation of hindering is that there is an initial, unconstrained
``natural'' rate of growth, but as $Q$ increases, its rate of growth is
constrained and tends to diminish, consistent with the notion of decreasing
marginal productivity. Based on the logistic, ecological models of population
growth invoke $r$- and $K$-selection \cite{Pianka70, Schacht80}, the respective
equivalents of unhindered and hindered growth. This terminology reflects the
notation for $r$ as the maximal intrinsic rate of natural increase (\gu\ in our
notation) and $K$ the carrying capacity. The concept of $r$- and $K$-selection
is a restricted application of the general formalism presented here. The
hindering formalism  is not limited to the logistic or any other growth
pattern; instead of the carrying capacity $K$, the impact of hindering is
characterized by the hindering parameter \Qh, whose definition (eq.\
\ref{eq:Qh}) is applicable to all patterns of decelerated growth.

A phenomenological description of data would not be particularly useful if it
involved an unwieldy number of parameters. However, all cases studied to date
required no more than two hindering terms \cite{EKZ20, Elitzur21}, indicating
that the hindering approach did capture essential properties of the growth
process in those cases. The role of successive hindering terms is clearly
visible in the fits of \cov\ cases in New York (Figure \ref{fig:NY_covid}). The
US population modeling, too, is instructive. Removing the hindering term from
the best-fitting model (\hbox{\S\ref{sec:US}}) turns it into a simple
exponential function. This exponential is a nearly perfect fit for the first 25
years of data, but applying it to the rest of the time series implies a 2020 US
population of almost 9 billion(!), more than 27 times the actual value. A
single $k = 1$ hindering term transforms this exponential into the model shown
in Figure \ref{fig:US}; the model result for the year 2020 is now 323 million,
within 2\% of the actual population. A successful correction of this magnitude
with just a single parameter is unlikely to be a mere coincidence.

{\new
\subsection{Limitations, challenges, future work}
}

The hindering formalism deals exclusively with long-term trends, ignoring the
fluctuations about trend lines. Its strength is not in reproducing details in
the data but in highlighting patterns of growth through analytic description
with the minimal number of free parameters. The simplicity and persistence of
long-term trends in the growth of US population and GDP uncovered by the
analysis (Figure \ref{fig:US}) is striking, especially in light of the massive
upheavals during the covered period which include two world wars, the Great
Depression and the transformation of the US economy from agrarian to industrial
and then technological. The absence of large fluctuations of the US population
about the fitted model stands out. The 231 data points deviate from the model
an average of just under 2.5\%; there is hardly any evidence for the waves of
immigration and major changes to immigration laws during that time span. This
smooth behavior may be partly attributable to the inherent stability of
hindered growth, which has $dg/dQ < -g/Q$ (eq.\ \ref{eq:stability}): when $Q$
rises above the underlying growth trajectory, the growth rate decreases and $Q$
is driven back toward the growth pattern, with the opposite happening if $Q$
declines below the long-term trend line. The GDP underlying pattern shows great
persistence as well. While the trauma of the Great Depression is clearly
discernible, afterward the GDP time variation reverts to the same simple
function that described earlier epochs. The GDP fluctuations are both large and
frequent, but subsided considerably after World War II: the average deviation
of model from data is 11.3\% before 1950 but only 3.6\% after. It appears that
government action had little effect in modifying the underlying growth pattern
of either US population or GDP but did have a significant impact on dampening
GDP fluctuations in recent years.

{\new

As this brief discussion shows, the hindering formalism is an ideal detrending
tool for time-series analysis when the long-term trend is one of growth. The US
GDP modeling results show that residuals, too, may contain important additional
structure that would require other data analysis methods. Integrating the
hindering formalism into the existing extensive framework of time-series
analysis is a major task for future work.
}

Hindering, the negative impact of a growing $Q$ on its own growth, is not the
only process that can cause growth-rate variations. Such variations can also
arise from changes to the environment in which $Q$ is growing. In the island
example (\S\ref{sec:hindering}), climate change could affect tree growth and
vary the inherent growth rate \gu. The processes driving growth-rate variation
are immaterial to our solution of the equation of growth (eq.\
\ref{eq:solution}). The basic premise of the solution procedure, that $g$ can
be considered a function of $Q$ instead of $t$, hinges on $Q$ being a
single-valued function of $t$, and this holds for every monotonically
increasing $Q$. However, hindering depends inherently on $Q$, reflecting
negative feedback to its environmental impact, while time variation of the
environment is inherently a function of $t$, unrelated to the growing $Q$.
While mathematically justified, expressing in terms of $Q$ a $t$-variation that
is inherently independent of $Q$ can be expected to increase complexity under
most circumstances. The simplicity of the models in Figures \ref{fig:US} and
\ref{fig:NY_covid} therefore suggests that hindering is the more plausible
driver of growth deceleration in these cases. Since the environment is
certainly varying, this indicates that significant changes to the environment
take longer than the hindering time scale, which can be taken as the doubling
time for $Q$,\footnote{Growth is described by the independent variable $x =
t/\Tu$, where $\Tu = 1/\gu$ is the growth time during the unhindered phase
(eq.\ \ref{eq:x}). The associated doubling time is $\Tu\ln2$, which is 21 years
for the US population, 18 years for the US GDP and 1.44 days(!) for the NY
\cov\ outburst.} enabling the growth pattern to adjust smoothly to the changing
environment. By contrast, environmental changes completed over periods shorter
than the doubling time are akin to phase transitions between states of matter
--- the system switches mid-growth to a state with different characteristics.
Preliminary work indicates that such abrupt transitions may be found in some
GDP and population data. This is an important topic for future studies.

The hindering formalism is based on a general mathematical solution of the
equation of growth and thus should be applicable in a wide variety of growth
situations, including, for example, biological and physical systems. Indeed,
the impetus for this work came from the growth of laser and maser
radiation,\footnote{The word laser is acronym for Light Amplification by
Stimulated Emission of Radiation. Similarly, masers involve Microwave instead
of Light. Requiring special conditions on earth, maser amplification occurs
naturally in many astronomical sources; a popular exposition is available in
\cite{Elitzur95}.} where growth equations are derived from first principles of
radiation theory that describe the dynamics of the underlying physical
processes \citep{Elitzur92}. In that case the growth pattern is $k = 1$ sth
(\S\ref{sec:sth}), with parameters derived from coefficients that describe
various aspects of fundamental interactions between matter and radiation.
{\new %
However, the solution is inapplicable when the growth rate becomes negative and
the time-series switches to a long-term trend of contraction instead of
expansion. A prominent example is the population of Japan, which according to
UN
data\footnote{\url{https://population.un.org/wpp/Download/Standard/Population/}}
has been in continuous decline since 2009. Declining trends present two
problems. The first is that the studied quantity is no longer a single-valued
function of time, thus the growth rate cannot be considered a function of $Q$
instead of $t$, the crucial first step in the general solution of the growth
equation \hbox{(\S\ref{sec:growth}).} This problem is a mere technicality,
though, and can be solved by dividing the time-series into segments, each with
a single-trend behavior.

The second, more serious problem is that the unhindered growth rate \gu, a
crucial ingredient of the hindering formalism, becomes meaningless for a
decreasing quantity. This ``natural'' growth rate, determined from the $Q \to
0$ limit of $g$ (eq.\ \ref{eq:gu}), is a fundamental property of the system
with an intrinsic, well defined meaning. Invoking again the island example
(\S\ref{sec:hindering}), in principle \gu\ could be determined even if apple
seeds were never actually introduced into the island. By contrast, a
contracting system does not offer an obvious intrinsic scale that does not
depend on initial conditions. Because of this fundamental difficulty, a general
description of negative growth situations requires a different approach and
remains an important challenge for future work.
}

\acknowledgments{I have greatly benefited from discussions with Joseph
Friedman, Željko Ivezić, Scott Kaplan, Dejan Vinković and David Zilberman.}

\bigskip
\appendixtitles{yes}
\appendixstart
\appendix

\section{The Gompertz Curve}
\label{app:Gompertz}

Employed often by demographers and actuaries to describe the distribution of
adult life spans \cite{Vaupel86, Willemse00}, the Gompertz function can be
written as \cite{Winsor32}

\eq{
   Q(t) = K e^{-b e^{-t/\tau}}
}
with $K$, $b$ and $\tau$ positive constants. Like the logistic, this function
has an upper bound, the carrying capacity $K = Q(t \to \infty)$. Its growth
rate as a function of $Q$

\eq{
    g(Q) = \frac{1}{\tau} \ln \frac{K}{Q}
}
has a singularity in the limit $Q \to 0$. Similarly, the growth rate as a
function of $t$

\eq{
    g(t) = \frac{b}{\tau} e^{-t/\tau}
}
diverges exponentially when $t \to -\infty$ (i.e., $Q \to 0$). As a result, the
unhindered growth rate (eq.\ \ref{eq:gu}) cannot be defined. The Gompertz
function cannot be incorporated into the general hindering formalism described
here.

%%%%%%%%%%%%%%%%%%%%%%%%%%%%%%%%%%%%%%%%%%
\end{paracol}
%%%%%%%%%%%%%%%%%%%%%%%%%%%%%%%%%%%%%%%%%%
\reftitle{References}

%\externalbibliography{yes}
%%\bibliography{../Econ}
%\bibliography{paper_mdpi_rev1}
%
%\end{document}

\end{document}